\documentclass[aps,prd,eqsecnum,showpacs,amsmath]{revtex4}
\usepackage{graphicx}
\begin{document}

\thispagestyle{empty}

\title{Stability criterion for self-similar solutions with a scalar
field and those with a stiff fluid in general relativity}
\author{
Tomohiro Harada $^{1}$\footnote{Electronic
 address:T.Harada@qmul.ac.uk} and
Hideki Maeda $^{2}$\footnote{Electronic
 address:hideki@gravity.phys.waseda.ac.jp}}
\affiliation{
$^{1}$Astronomy Unit, School of Mathematical Sciences,
Queen Mary, University of London,
Mile End Road, London E1 4NS, UK\\
$^{2}$Advanced Research Institute for Science and Engineering,
Waseda University, Tokyo 169-8555, Japan
}
\date{\today}

\begin{abstract}                
A stability criterion is derived in general relativity 
for self-similar solutions with 
a scalar field and those with a stiff fluid, which is a perfect 
fluid with the equation of state $P=\rho$.
A wide class of self-similar solutions turn out to be unstable 
against kink mode perturbation.
According to the criterion, 
the Evans-Coleman stiff-fluid solution is unstable and cannot be 
a critical solution for the spherical collapse of a stiff fluid
if we allow sufficiently small 
discontinuity in the density gradient field
in the initial data sets.
The self-similar scalar-field solution, which was recently
found numerically by Brady {\it et al.} 
(2002 {\it Class. Quantum. Grav.} {\bf 19} 6359), is also unstable.
Both the flat Friedmann universe
with a scalar field and that with a stiff fluid 
suffer from kink instability at the particle
horizon scale.  
\end{abstract}

\maketitle

\section{Introduction}
Since the Einstein equations are simultaneous nonlinear
partial differential equations,
it is not easy to obtain general solutions.
In this context,
we can assume {\em self-similarity}
and consider {\em self-similar solutions}.
For example, in spherical symmetry, 
this assumption reduces the partial differential
equations to a set of 
ordinary differential equations~\cite{ct1971}
and even the classification of self-similar 
solutions with a perfect fluid has been  
done~\cite{gnu1998,cc2000,ccgnu2000} 
based on the theory of dynamical systems (\cite{we1997,coley1999} and 
references therein).
Thus, the assumption of self-similarity is 
very powerful in finding
dynamical and inhomogeneous solutions. 
The application of self-similar solutions 
is very large, including cosmological perturbations,
star formation,
gravitational collapse, 
primordial black holes~\cite{ch1974,lcf1976,bh1978},
cosmological voids and cosmic censorship~\cite{op1987,op1990,fh1993}.
See~\cite{cc1999} for a brief review of 
self-similar solutions.

In recent progress in general relativity,
self-similar solutions have attracted attention 
not only because they are easy to obtain 
but also because they may play important roles 
in cosmological situations and/or gravitational collapse.
One of the important roles of self-similar solutions 
is to describe asymptotic behaviours of more 
general solutions.
This feature has been found in several models
such as homogeneous cosmological models~\cite{we1997,coley1999}.
In a cosmological context, the suggestion that 
spherically symmetric fluctuations might naturally evolve
from complex initial conditions 
to a self-similar form has been termed the {\em similarity
hypothesis}~\cite{carr1993,cc1999}.
The hypothesis has been also suggested to hold in more 
general situations, including gravitational collapse.
In fact, Harada and Maeda~\cite{hm2001} numerically found that
a self-similar solution acts as an {\em attractor} 
in the spherically symmetric collapse of a perfect fluid with
the equation of state $P=k\rho$ ($0<k<1$) 
at least for $0<k\alt 0.03$ (see also~\cite{harada1998}).

Spherically symmetric self-similar solutions 
have been also studied in the context of 
{\em critical behaviour} in gravitational collapse, in which
a self-similar solution is not an attractor but an 
{\em intermediate attractor}.
The critical behaviour was numerically discovered 
by Choptuik~\cite{choptuik1993} 
in the spherical system of a massless scalar field.
There appears a {\em discretely self-similar solution}
at the threshold of black hole formation for $p=p^{*}$,
where $p$ and $p^{*}$ are the parameter of initial data sets
and the critical value of the parameter, respectively.
For supercritical collapse, 
the black hole mass $M_{\rm BH}$ follows the scaling law
$M_{\rm BH}\propto |p-p^{*}|^{\gamma}$,
where $\gamma\approx 0.37 $ is referred to as
the {\em critical exponent}.
Subsequently,
these findings have been confirmed by several 
authors (e.g.~\cite{hs1996}).
Evans and Coleman~\cite{ec1994} studied the spherical system of 
a radiation fluid ($P=\rho/3$) and found similar phenomena
although the observed critical solution is 
a {\em continuously self-similar solution}.
Koike, Hara and Adachi~\cite{kha1995} showed that the critical behaviour 
is understood by 
the intermediate behaviour around a saddle equilibrium point. 
This approach was also successfully 
applied to 
the spherical system of a massless scalar field to understand
the critical behaviour observed by Choptuik~\cite{gundlach19951997}.
Several 
authors~\cite{maison1996kha1999,nc2000,hm2001} 
have extended the work to
perfect fluids with the equation of state $P=k\rho$ ($0<k \le 1$).
The limiting case $k\to 0$ turns out 
to be the Newtonian critical phenomena~\cite{mh2001}.
Another case $k=1$, for which the perfect fluid 
is called a {\em stiff fluid}, is interesting because the stiff
fluid is known to be equivalent to a massless scalar field in 
certain circumstances~\cite{madsen1988cg1999,bcgn2002}.
Neilsen and Choptuik~\cite{nc2000} studied the stiff-fluid system and 
found that the critical behaviour can be regarded as
a continuous limit of that for a perfect fluid
with $P=k\rho$ ($0<k<1$) as $k\to 1$. 
The critical solution is continuously self-similar and 
the critical exponent is given by $\gamma=0.92\pm0.02$.
Brady {\it et al.}~\cite{bcgn2002} studied the mystery of why the critical
behaviour for a stiff fluid is different from that for a scalar field.
See~\cite{gundlach2003} for a broad review of critical phenomena
in gravitational collapse.

The present article concerns {\em weak discontinuity} 
around self-similar solutions.
It has been known that the required differentiability condition
for the spacetime metric tensor can be much weaker than $C^{\infty}$,
for the existence and uniqueness of geodesics and/or 
well-defined initial value formulation of the Einstein 
equation~\cite{he1973}.
Moreover, it has been well established that 
shock waves, which is one of the examples of strong discontinuity, 
play essential roles in many astrophysical situations. 
If we insert a perturbation with a sufficiently small
weak discontinuity,
it may grow too large to neglect as time proceeds.
This instability has been called {\em kink instability}.
The kink instability of self-similar solutions
was studied by Ori and Piran~\cite{op1988} for the spherical system
of isothermal gas in Newtonian gravity.
Harada~\cite{harada2001} (paper I) investigated the kink instability 
of self-similar solutions for the spherical system of 
a perfect fluid with the equation of state $P=k\rho$ 
($0<k<1$) in general relativity.
In this article we derive a stability criterion against
the kink mode perturbation for both self-similar solutions with
a scalar field and those with a stiff fluid in general 
relativity. 
Several interesting applications are discussed,
one of which is related to 
critical behaviour for these two
systems. 

This article is organized as follows. In Sec.~II, we derive the 
basic equations for the spherically symmetric Einstein-scalar-field system
and those for self-similar solutions.
Also in this section we introduce the kink mode for 
the self-similar scalar-field 
solutions, derive the equations for this mode and obtain the 
stability criterion for self-similar scalar-field solutions.
In Sec.~III, the analysis in paper I is extended to include 
the stiff-fluid system.
Section~IV is devoted to obtain the correspondence relation 
between self-similar 
solutions of the two systems, the scalar-field system and 
the stiff-fluid one.
In Sec.~V, we discuss the applications of the present result,
in particular, to critical behaviour of gravitational collapse.
In Sec.~VI, we summarize the paper.
We adopt units such that $G=c=1$ and the abstract 
index notation of~\cite{wald1983}.

\section{Stability criterion for self-similar solutions with a scalar field}
\subsection{Equations for a self-gravitating massless scalar field}
In this section, we basically follow the notation
given by Brady {\it et al.}~\cite{bcgn2002}. 
We consider a massless scalar field $\phi$ as a matter field,
for which the stress-energy tensor is given by 
\begin{equation}
T^{ab}=\nabla^{a}\phi\nabla^{b}\phi-\frac{1}{2}g^{ab}\left(\nabla_{c}\phi
\nabla^{c}\phi\right).
\label{eq:stress-energy_tensor_of_scalar_field}
\end{equation}
We adopt the Bondi coordinates for a spherically symmetric 
spacetime as
\begin{equation}
ds^{2}=-g\bar{g}du^{2}-2gdudR+R^{2}(d\theta^{2}+\sin^{2}\theta d\phi^{2}),
\label{eq:Bondi_coordinates}
\end{equation}
where $g=g(u,R)$ and $\bar{g}=\bar{g}(u,R)$.
Then the Einstein equation and the equation of motion for the scalar field
reduce to the following partial differential equations:
\begin{eqnarray}
(\ln g)_{,R}&=&4\pi R \phi_{,R}^2,\label{pbasic1}\\
R {\bar g}_{,R}&=&g-{\bar g},\label{pbasic2}\\
g\left(\frac{{\bar g}}{g}\right)_{,u}&=&8\pi R(\phi_{,u}^2-{\bar g}\phi_{,u} \phi_{,R}),\label{pbasic3}\\
(R^2 {\bar g}\phi_{,R})_{,R}&=&2R\phi_{,u}+2R^2\phi_{,uR},\label{pbasic4}
\end{eqnarray}
where the comma denotes the partial derivative.
For later convenience we define a new function ${\bar h(u,R)}$ as
\begin{eqnarray}
\phi={\bar h(u,R)} -\kappa\ln|u|,
\end{eqnarray}
where $\kappa$ is an arbitrary constant, and we define the following variables:
\begin{equation}
x \equiv  -\frac{R}{u}, \quad X\equiv \ln|x|,\quad T \equiv  -\ln |u|.
\label{eq:xXT}
\end{equation}
Then equations (\ref{pbasic1})--(\ref{pbasic4}) reduce to 
\begin{eqnarray}
(\ln g)'&=&4\pi {\bar h}'^2,\label{basic1}\\
{\bar g}'&=&g-{\bar g},\label{basic2}\\
g\left[\left(\frac{{\bar g}}{g}\right)^{\dot{}}+\left(\frac{{\bar g}}{g}
\right)'\right]
&=&8\pi ({\dot {\bar h}}+{\bar h}'+\kappa)[x({\dot {\bar h}}+{\bar h}'+\kappa)-{\bar g}{\bar h}'],\label{basic3}\\
({\bar g}{\bar h}')'+{\bar g}{\bar h}'&=&2x[({\dot {\bar h}}+{\bar h}')+({\dot {\bar h}}+{\bar h}')'+\kappa],\label{basic4}
\end{eqnarray}
where the dot and prime denote the partial derivatives with respect to $T$
and $X$, respectively. 
We refer to $u<0$ and $u>0$ as early times and 
late times, respectively.
We refer to $u<0$ and $u>0$ as early times and late times, respectively.
It is noted that the present definition of prime
is different from that in Brady {\it et al.}~\cite{bcgn2002}.

\subsection{Self-similar scalar-field solutions}
\label{subsec:self-similar_scalar-field_solutions}
For self-similar solutions, we assume that 
$g=g(X)$, ${\bar g}={\bar g}(X)$ and ${\bar h}={\bar h}(X)$. 
The ordinary differential equations for $g$, $\bar{g}$ and $\bar{h}$
are given by
\begin{eqnarray}
(\ln g)'&=& 4\pi j^{2}, 
\label{eq:g'}\\
\bar{g}'&=& g-\bar{g}, 
\label{eq:gbar'}\\
g-\bar{g}&=&4\pi [2\kappa^{2}x-(\bar{g}-2x)(j^{2}+2\kappa j)], \\
\bar{h}'&=&j, 
\label{eq:g-gbar}\\
(2x-\bar{g})j'&=&-2\kappa x+j(g-2x),
\label{eq:j'}
\end{eqnarray}
where the prime denotes the differentiation 
with respect to $X$. 

These equations are singular when $\bar{g}=2x$,
which is called a {\em similarity horizon}.
We denote the value of $x$ at the similarity horizon as
$x_{\rm s}$ and also the value of $X$ as 
$X_{s}\equiv \ln |x_{s}|$.
This line corresponds to a radial ingoing (outgoing) null curve 
for early-time (late-time) solutions because 
\begin{equation}
\frac{dX}{dT}=1-\frac{\bar{g}}{2x},
\end{equation}
along a radial ingoing (outgoing) null curve.
We consider self-similar scalar-field solutions 
with finite values of 
functions $g$, $\bar{g}$, $\bar{h}$ and $j$ 
and their gradients with respect to $X$ at the similarity horizon.
Then we find at $X=X_{\rm s}$
\begin{equation}
g = 2x_{\rm s}(4\pi\kappa^{2}+1), \quad
\bar{g} = 2x_{\rm s}, \quad
j = \frac{1}{4\pi\kappa},
\label{eq:regularity_similarity_horizon}
\end{equation}
for $\kappa\ne 0$ and 
\begin{equation}
g=\bar{g}=2x_{\rm s}, \quad
j = j_{\rm s},
\label{eq:regularity_similarity_horizon_kappa0}
\end{equation}
for $\kappa=0$. For $\kappa=0$, similarity horizons are 
parametrized by $j_{\rm s}$.
Since we have
\begin{equation}
\left(1-\frac{\bar{g}}{2x}\right)'=1-4\pi\kappa^{2},
\end{equation}
at the similarity horizon,
the self-similarity horizon is ``transluminal'' , i.e. 
it has subluminal interior and superluminal exterior for $4\pi \kappa^{2}<1$,
while it is ``anti-transluminal'', i.e.
it has superluminal interior and subluminal exterior for $4\pi \kappa^{2}>1$.

We can have regular centre at $R=0$ for $u\ne 0$, which corresponds to 
$X=-\infty$.
The regularity condition reduces to 
the following initial condition at $X=-\infty$.
\begin{equation}
g=\bar{g}=g_{0}, \quad
\bar{h}=\bar{h}_{0}, \quad
j=0.
\label{eq:gbar0_g0_h0_j0}
\end{equation}
Taking the limit $X\to -\infty$ in Eq.~(\ref{eq:j'}) and
using the l'Hospital's theorem, we have
\begin{equation}
\lim_{X\to -\infty}\frac{j'(X)}{x(X)}
=\lim_{X\to -\infty}\frac{j(X)}{x(X)}=\frac{\kappa}{g_{0}}.
\label{eq:j'zero}
\end{equation}

The above system of Eqs.~(\ref{eq:g'})--(\ref{eq:j'}) was
studied by 
Christodoulou~\cite{christodoulou1994}, 
Brady~\cite{brady1995} and 
Brady {\it et al.}~\cite{bcgn2002}.
Exact self-similar scalar-field solutions discovered by 
Roberts~\cite{roberts1989} correspond to general solutions for
$\kappa=0$,
which will be described in some detail in Appendix~\ref{sec:roberts_solution}.
Christodoulou~\cite{christodoulou1994} found that there is a solution with
$4\pi\kappa^{2}=1/3$, which has both regular centre and analytic 
similarity horizon and turns out to be the flat Friedmann solution. 
Using a two-point shooting method, Brady {\it et al.}~\cite{bcgn2002}
numerically found another self-similar scalar-field solution with
both regular centre and analytic similarity horizon,
which we will discuss in Secs.~\ref{subsec:Evans-Coleman} 
and \ref{subsec:BCGN} in detail.
\subsection{Kink instability of self-similar solutions with a scalar field}
\label{subsec:kink_instability_scalar_field}
We consider perturbations which satisfy the following conditions in the background of a self-similar solution:
(1) The initial perturbations vanish inside the similarity horizon
for early-time solutions ($u<0$). 
(For late-time solutions ($u>0$), 
the initial perturbations vanish outside the similarity horizon.)
(2) $g$, ${\bar g}$, ${\bar h}$ and ${\bar h}'$ are continuous everywhere, in particular at the similarity horizon.
(3) ${\bar h}''$ and $\dot{\bar h}''$ are discontinuous at the
similarity horizon, although they have finite one-sided limit values as 
$X\to X_{\rm s}-0$ and $X\to X_{\rm s}+0$.

We denote the full-order perturbations as
\begin{equation}
\delta g(T,X)\equiv g(T,X)-g_{\rm b}(X),\quad
\delta {\bar g}(T,X)\equiv {\bar g}(T,X)-{\bar g}_{\rm b}(X), \quad
\delta {\bar h}(T,X)\equiv {\bar h}(T,X)-{\bar h}_{\rm b}(X), \quad
\end{equation}
where $g_{\rm b}$, ${\bar g}_{\rm b}$ and ${\bar h}_{\rm b}$ 
denote the background self-similar solution.
 
By conditions (2), the perturbations satisfy 
$\delta {\bar h}= \delta {\bar h}'=0$ and $\delta {\bar h}''\ne 0$ 
at the similarity horizon at 
initial moment $T=T_{0}$. 
The evolution of the initially unperturbed 
region is completely described by the background 
self-similar solution because no information from the perturbed side can
penetrate the unperturbed side by condition (1). 
Then, by conditions (2) and (3), we find 
$\delta {\bar h}=\delta {\bar h}'=0$ and $\delta {\bar h}'' \ne 0$ 
at the similarity horizon for 
$T\ge T_{0}$ for early-time solutions 
(for $T\le T_{0}$ for late-time solutions). 
We find $\delta g=\delta \bar{g}=\delta g'=
\delta {\bar g}'=\delta {\bar g}''=0$ but 
$\delta g''\ne 0$ and $\delta \bar{g}'''\ne 0$ 
from Eqs.~(\ref{basic1}), (\ref{basic2}) and (\ref{basic3}) 
for the full-order perturbations at the point of discontinuity.

Differentiating Eq.~(\ref{basic1}) with respect to $X$ 
and estimating both sides at the point of discontinuity, we obtain 
\begin{equation}
\delta g''=8\pi g_{\rm b}j_{\rm b}\delta j'.
\end{equation}
Using Eqs.~(\ref{basic1}) and (\ref{basic2}), 
differentiating Eq.~(\ref{basic3}) with respect to $X$, 
and estimating both sides at the point of discontinuity, we obtain
\begin{equation}
(j_{\rm b}+\kappa)({\bar g}_{\rm b}-2x)\delta j'=0. 
\label{discontinuity=similarity_horizon}
\end{equation}
Therefore, $\delta j'\ne 0$ is possible at the similarity horizon
$X=X_{\rm s}$ where ${\bar g}_{\rm b}=2x_{\rm s}$.

Using Eqs.~(\ref{basic1}), (\ref{basic2}) and
(\ref{discontinuity=similarity_horizon}), 
differentiating Eq.~(\ref{basic3}) twice with respect to $X$, 
and estimating both sides at the point of discontinuity, we obtain
\begin{equation}
2x(j_{\rm b}+\kappa)\delta \dot{j}'-[g_{\rm b}j_{\rm b}+2(g_{\rm b}-4x)(j_{\rm b}+\kappa)] \delta j'=(j_{\rm b}+\kappa)(\bar{g}_{\rm b}-2x)\delta j''.
\end{equation}
We can have another equation by 
using Eq.~(\ref{basic2}), 
differentiating Eq.~(\ref{basic4}) with respect to $X$, 
and estimating both sides at the point of discontinuity as
\begin{equation}
2x \delta \dot{j}'-(2g_{\rm b}-\bar{g}_{\rm b}-4x)\delta j'=(\bar{g}_{\rm b}-2x) \delta j''.
\end{equation}
The two equations obtained above are independent in general but
degenerate at the similarity horizon. At the similarity horizon, 
we find that these two equations reduce to 
\begin{equation}
\delta\dot{j}'-(8\pi \kappa^{2}-1)\delta j'=\lim_{X\to X_{\rm s}}\left[\left(\frac{\bar{g}_{\rm b}}{2x}-1\right)\delta j''\right],
\label{eq:higher_derivative_scalar_field}
\end{equation}
where Eq.~(\ref{eq:regularity_similarity_horizon}) or
Eq.~(\ref{eq:regularity_similarity_horizon_kappa0}) is used.
Since the left-hand side has a finite limit at the similarity horizon,
the right-hand side also has a finite limit.
Since $\delta j''$ may not have a finite limit at the similarity
horizon, the limit value on the right-hand side 
is not trivial. However, we can show that the 
right-hand side vanishes employing the 
proof in Appendix~\ref{sec:higher_derivative}.
Therefore, we obtain the full-order 
perturbation equation for $\delta j'$ at the similarity horizon:
\begin{equation}
\delta \dot{j}'-(8\pi \kappa^{2}-1)\delta j'=0.
\label{eq:kink_mode_equation_scalar_field}
\end{equation}
It should be noticed that the full-order analysis results in 
a linear-order equation.
This equation can be integrated as
\begin{equation}
\delta j'=\mbox{const}\cdot e^{\alpha T},
\label{eq:kink_mode_solution_for_scalar_field}
\end{equation}
where 
\begin{equation}
\alpha=-(1-8\pi \kappa^{2}).
\label{eq:growth_rate_for_scalar_field}
\end{equation}
Therefore, for early-time solutions, it is found that
the perturbation decays exponentially for $4 \pi \kappa^2 < 1/2$, 
it is constant for $4 \pi \kappa^2 = 1/2$ 
and it grows exponentially for $4 \pi \kappa^2 > 1/2$.
The situation is reversed for late-time solutions.

Here we define instability by the exponential growth of discontinuity.
Then we find the following criterion: for early-time solutions,
solutions with a regular similarity horizon and $4\pi \kappa^{2}<1/2$
are stable against the kink mode, while those with $4\pi \kappa^{2}>1/2$
are unstable. Solutions with $4\pi \kappa^{2}=1/2$ are marginally stable
against this mode. The situation is reversed for late-time
solutions.
\section{Stability criterion for self-similar solutions with a stiff fluid} 

\subsection{Equations for a self-gravitating stiff fluid}
In Sec.~\ref{subsec:kink_instability_scalar_field}
we have seen that the kink instability can occur
in self-similar scalar-field solutions.
Since a scalar field is equivalent to a stiff fluid,
it is expected that
self-similar stiff-fluid solutions can also have 
the kink instability.
However, the equivalence turns out to be incomplete
as we will see in Sec.~\ref{sec:correspondence}.
Hence it will be interesting to see when and how the kink instability
occurs in stiff-fluid self-similar solutions. 
In paper I, 
the kink instability is studied in perfect-fluid systems
with the equation of state $P=k\rho$ ($0<k<1$).
Almost all equations in paper I are available also 
for a stiff fluid with $k=1$.
Here we avoid repetition of the derivation 
and only quote the equations which are
necessary for the subsequent analysis.
We basically follow the formulation given by~\cite{bh1978}.

The stress-energy tensor for a stiff fluid is given by
\begin{equation}
T^{ab}=\rho(2 u^{a}u^{b}+g^{ab}),
\label{eq:stress-energy_tensor_of_stiff_fluid}
\end{equation}
where $\rho$ is the energy density and 
$u^{a}$ is the four velocity of the fluid element.
In a spherically symmetric spacetime, there is a natural quasilocal
mass called the Misner-Sharp mass, defined by~\cite{ms1964}
\begin{equation}
m\equiv \frac{R}{2}\left(1-\nabla^{a}R\nabla_{a}R\right),
\end{equation} 
where $R$ is the circumferential radius.
The line element in a spherically symmetric spacetime is given by
\begin{equation}
ds^{2}=-e^{\sigma(t,r)}dt^{2}+e^{\omega(t,r)}dr^{2}+R^{2}(t,r)(d\theta^{2}
+\sin^{2}\theta d\phi^{2}),
\label{eq:diagonal}
\end{equation}
where we choose the radial coordinate $r$ to be 
comoving with fluid elements. 
We define dimensionless functions such as
\begin{equation}
\eta \equiv  8\pi r^{2}\rho,\quad
S \equiv \frac{R}{r}, \quad
M \equiv  \frac{2m}{r}, \quad
y\equiv\frac{M}{\eta S^{3}},
\end{equation}
where we have also defined an auxiliary function $y$ for later 
convenience.
We introduce the following variables:
\begin{equation}
\xi\equiv  \frac{t}{r}, \quad
\tau  \equiv   -\ln |t|, \quad
z  \equiv  -\ln|\xi|.
\end{equation}
The velocity function $V$ is defined by the velocity of the 
$\xi=\mbox{const}$ curve relative to the fluid element,
which is written as
\begin{equation} 
V= \mbox{sign}(\xi)e^{z+\frac{\omega-\sigma}{2}}.
\end{equation}

We can find that the equations of motion for the fluid are integrated as
\begin{equation}
e^{\sigma}=a_{\sigma}
\eta^{-1} e^{2z}, \quad e^{\omega}=a_{\omega}\eta^{-1}S^{-4},
\label{eq:esigmaeomega}
\end{equation}
where we make $a_{\sigma}$ and $a_{\omega}$ 
constant using coordinate transformation.
Using Eqs.~(\ref{eq:esigmaeomega}), $V^{2}$ is rewritten as
\begin{equation}
V^{2}=\frac{a_{\omega}}{a_{\sigma}}S^{-4}.
\label{eq:V2}
\end{equation}
Using the dimensionless variables, 
we can derive the following equations from the Einstein equations:
\begin{eqnarray}
& &M+M'=\eta S^{2}(S+S'), 
\label{eq:tauz1}\\
& &\dot{M}+M'= -\eta S^{2}(\dot{S}+S'), 
\label{eq:tauz2}\\
& &\ddot{S}+2\dot{S}'+S''+\left[\frac{1}{2}\left(\frac{\dot{\eta}}
{\eta}+\frac{\eta'}{\eta}-2\right)+1\right](\dot{S}+S')\nonumber \\
\label{eq:tauz3}
&=&-\frac{1}{2}
e^{\sigma-\omega-2z}\left(\frac{\eta'}{\eta}-2\right)
(S+S')-\frac{1}{2}e^{\sigma-2z}
\frac{M+\eta S^{3}}{S^{2}},
\label{eq:tauz4}
\end{eqnarray}
and 
\begin{equation}
\frac{M}{S}=1+e^{-\sigma+2z}(\dot{S}+S')^{2}-e^{-\omega}(S+S')^{2}, \\
\label{eq:tauz5}
\end{equation}
where $e^{\sigma}$ and $e^{\omega}$ are given by 
Eq.~(\ref{eq:esigmaeomega}), 
the dot and prime denote the partial derivatives with respect to
$\tau$ and $z$, respectively, and 
three of the above four equations are independent.

\subsection{Self-similar stiff-fluid solutions}
\label{subsec:self-similar_stiff-fluid_solutions}
For self-similar solutions, we assume that all dimensionless
quantities depend only
on $z$, i.e., 
$M=M(z)$, 
$S=S(z)$,
$\eta=\eta(z)$, 
$\sigma=\sigma(z)$, and 
$\omega=\omega(z)$.
Then, from equations~(\ref{eq:tauz1})--(\ref{eq:tauz5}), 
we obtain the following ordinary differential equations:
\begin{eqnarray}
M'&=&\frac{1}{2}\frac{1-y}{y}M, 
\label{eq:M'} \\
S'&=&-\frac{1-y}{2}S, 
\label{eq:S'}\\
\eta'&=&2\left[(1-y)+\frac{y-a_{\sigma}V^{2}}
{1-V^{2}}\right]\eta,
\label{eq:eta'}
\end{eqnarray}
and constraint equation
\begin{equation}
V^{2}(1-y)^{2}-(1+y)^{2}+4
a_{\sigma}V^{2}y
\left(\frac{S}{M}-1\right)=0,
\label{eq:const}
\end{equation}
where $V^{2}$ is given by Eq.~(\ref{eq:V2}).
It should be noted that $a_{\sigma}$ does not change under the 
scale transformation and 
therefore its value has a physical meaning. $a_{\sigma}$ is 
physically characterized
by the following relation:
\begin{equation}
a_{\sigma}=e^{\sigma}\eta e^{-2z}=8\pi \rho e^{\sigma} t^{2}.
\end{equation}

A {\em sonic point} $z=z_{\rm s}$ is defined by 
\begin{equation}
V^{2}=1.
\label{eq:sonicpoint1}
\end{equation}
Since the sound speed is equal to the light speed, 
the sonic point may correspond to the Cauchy horizon,
event horizon or particle horizon, simultaneously.
From Eqs.~(\ref{eq:V2}), (\ref{eq:const}) and
(\ref{eq:sonicpoint1}), we find $M=M_{\rm s}$ and $S=S_{\rm s}$
at the sonic point, where
\begin{equation}
M_{\rm s}= \frac{a_{\sigma}}{1+a_{\sigma}}
\left(\frac{a_{\omega}}{a_{\sigma}}\right)^{1/4}, \quad
S_{\rm s}= \left(\frac{a_{\omega}}{a_{\sigma}}\right)^{1/4}.
\end{equation}
Here we adopt the finiteness of the density gradient
as a regularity condition.
Then, from Eq.~(\ref{eq:eta'}), we obtain $y=y_{\rm s}$
at the sonic point, where 
\begin{equation}
y_{\rm s}= a_{\sigma}.
\label{eq:sonicpoint2}
\end{equation}

We introduce a new independent variable $u$, which is defined as
\begin{equation}
\frac{d z}{d u}=1-V^{2}.
\end{equation} 
Using $u$ in place of $z$, 
we have the following 
system of the ordinary differential equations:
\begin{eqnarray}
\frac{dM}{du}&=&\frac{1}{2}\frac{1-y}{y}(1-V^{2})M, \\
\frac{dS}{du}&=&-\frac{1-y}{2}(1-V^{2})S, \\
\frac{d\eta}{du}&=&2\left[(1-y)(1-V^{2})+
\left(y-a_{\sigma}V^{2}\right)\right]\eta.
\end{eqnarray}
In the above, the sonic point is found to be
an equilibrium point of the dynamical system
with four dimensional state space $\{(M,S,\eta,z)\}$.
More precisely, the whole of these
sonic points composes the two dimensional 
surface of equilibrium points determined by 
Eqs.~(\ref{eq:sonicpoint1}) and (\ref{eq:sonicpoint2}),
which is called a {\em sonic surface}. 
The qualitative properties of solutions are 
analyzed by the linearisation of the ordinary 
differential equations at the equilibrium point.
We put 
\begin{equation}
M=\eta_{\rm s}S_{\rm s}^{3}(y_{\rm s}+x_{1}), \quad 
S=S_{\rm s}(1+x_{2}), \quad
\eta=\eta_{\rm s}(1+x_{3}), \quad
\xi=\xi_{\rm s}(1+x_{4}) \quad (\mbox{or} \quad z=z_{s}-x_{4}),
\end{equation}
where $x_{1}$ to $x_{4}$ are
regarded as components of the vector 
${\bf x}$.
Then, we find the following linearised 
ordinary differential equations in the matrix form:
\begin{equation}
\frac{d}{du}{\bf x}={\bf A}{\bf x},
\end{equation}
where 
the matrix ${\bf A}$ is given by
\begin{equation}
{\bf A}=2\left(\begin{array}{cccc}
0 & 1-y_{\rm s} & 0 & 0 \\
0 & -(1-y_{\rm s}) & 0 & 0 \\
1 & (4-3y_{\rm s}) & -y_{\rm s} & 0 \\
0 & -2 & 0 & 0 
\end{array}\right).
\end{equation}
This matrix has two zero eigenvalues and two 
generically nonzero eigenvalues.
The former two zero eigenvalues are due to 
the two dimensional surface of equilibrium points
in four dimensional state space.
The latter two eigenvalues are:
\begin{equation}
\lambda_{1}=-2y_{\rm s},\quad\mbox{and}\quad \lambda_{2}=-2(1-y_{\rm s}),
\end{equation}
associated with the following eigenvectors:
\begin{equation}
\bf{e}_{1}\equiv \left(
\begin{array}{c}
0\\0\\1\\0
\end{array}
\right),\quad\mbox{and}\quad
\bf{e}_{2}\equiv \left(
\begin{array}{c}
(1-2y_{\rm s})(1-y_{\rm s})
\\-(1-2y_{\rm s})(1-y_{\rm s})\\3(1-y_{\rm s})^{2}\\-2(1-2y_{\rm s})
\end{array}
\right),
\end{equation}
respectively. Therefore, as we restrict our attention to 
the plane spanned by ${\bf e}_{1}$ and ${\bf e}_{2}$,
the sonic point is a node with primary direction
$\bf{e}_{1}$ and secondary direction $\bf{e}_{2}$ for $0<y_{\rm s}<1/2$,
a degenerate node for $y_{\rm s}=1/2$, 
a node with primary direction $\bf{e}_{2}$ and 
secondary direction $\bf{e}_{1}$ for $1/2<y_{\rm s}<1$,
and a saddle with repulsive-eigenvalue direction $\bf{e}_{2}$
and attractive-eigenvalue direction $\bf{e}_{1}$ for $1<y_{\rm s}$.

Both eigenvectors are regular in the whole four dimensional 
state space $\{(M,S,\eta,z) \}$ but not in the physical space.
Actually, along eigenvector $\bf{e}_{1}$, the density gradient
takes the following limit:
\begin{equation}
\lim_{z\to z_{\rm s}}\frac{\eta'}{\eta}=\pm\infty.
\end{equation}
The above divergence directly implies the divergence of 
the density gradient with respect to the physical length $l$
on the $t=\mbox{const}$ hypersurface as seen in the following 
equation:
\begin{equation}
\left(\frac{\partial \rho}{\partial l}\right)_{t}
=e^{-\omega/2}\left(\frac{\partial}{\partial r}
\frac{\eta}{8\pi r^{2}}\right)_{t}=e^{-\omega/2}\frac{\eta}{8\pi r^{3}}
\left(\frac{\eta'}{\eta}-2\right).
\end{equation}
Hereafter, we restrict our attention to the self-similar solutions
with finite density gradients at the sonic point. 
Under this restriction it is found that 
only the solutions along $\bf{e}_{2}$ are allowed
and that 
there is no acceptable solution with $y_{\rm s}=1/2$.
This is quite different from the case for $0<k<1$,
in which both directions associated with the two eigenvalues 
are physically acceptable. 
Along direction $\bf{e}_{2}$, we have
\begin{equation}
\frac{S'}{S}=-\frac{1}{2}(1-y_{\rm s}),
\end{equation}
at the sonic point for $y_{\rm s}\ne 1/2$.
Then, from Eq.~(\ref{eq:V2}), we have
\begin{equation}
(V^{2})'=2(1-y_{\rm s}),
\end{equation}
at the sonic point.
Therefore, the sonic point is ``transsonic''
for $0<y_{\rm s}<1/2$ or $1/2<y_{\rm s}<1$, 
while the sonic point is ``anti-transsonic''
for $1<y_{\rm s}$.

The classification is as follows.
For $0<y_{\rm s}<1/2$,
the solution crosses a nodal sonic point along
a secondary direction.
For $y_{\rm s}=1/2$,
the solution crosses a degenerate-nodal sonic point.
For $1/2<y_{\rm s}<1$,
the solution crosses a nodal sonic point along
a primary direction.
For $1<y_{\rm s}$, 
the solution crosses a saddle sonic point along
a repulsive-eigenvalue direction.
As for the family of acceptable solutions, 
for $0<y_{\rm s}<1/2$ or $1<y_{\rm s}$ 
there is only one acceptable solution which crosses
the sonic point and it has analyticity there
(see paper I for the analyticity of self-similar 
solutions at the sonic point.~\cite{footnote3})
For $1/2<y_{\rm s}\le 1$, there is a one-parameter family of acceptable
solutions which crosses the sonic point and only one of them
has analyticity there.
For $y_{\rm s}=1/2$, there is no acceptable solution.
It is clear that the case for $y_{\rm s}=1$ needs more careful
treatment because the equilibrium point is nonhyperbolic 
(c.f. the Hartman-Grobman theorem).
Bicknell and Henriksen~\cite{bh1978} found that there is a one-parameter
family of regular solutions in this case by a direct integration
around the sonic point.

\subsection{Kink instability of self-similar solutions with a stiff fluid}
\label{subsec:kink_instability_stiff_fluid}
We consider perturbations which satisfy
the following conditions in the background of a self-similar solution:
(1) The initial perturbations vanish inside the sonic point
for early-time solutions ($t<0$).
(The initial perturbations vanish 
outside the sonic point for late-time solutions ($t>0$).)
(2) $M$, $S$ and $\eta$ are continuous everywhere, 
in particular at the sonic point.
(3) $\eta^{\prime}$ and $\dot{\eta}'$ are discontinuous at the sonic point, 
although they have finite one-sided limit values as
$z\to z_{\rm s}-0$ and $z\to z_{\rm s}+0$~\cite{footnote1}.

We denote the full-order perturbations as
\begin{equation}
\delta S(\tau,z)\equiv S(\tau,z)-S_{\rm b}(z), \quad
\delta M(\tau,z)\equiv M(\tau,z)-M_{\rm b}(z), \quad
\delta \eta(\tau,z)\equiv \eta(\tau,z)-\eta_{\rm b}(z),
\end{equation}
where $M_{\rm b}$, $S_{\rm b}$ and $\eta_{\rm b}$ denote the background
self-similar solution.

For the perfect fluid case for the equation of state $P=k\rho$,
the evolution equation for the discontinuity at the sonic point 
becomes (paper I but~\cite{footnote2})
\begin{equation}
\frac{\delta \dot{\eta'}}{\eta_{\rm b}}+\left(
-2\frac{1-k}{1+k}\frac{\eta'_{\rm b}}{\eta_{\rm b}}+\frac{5-3k-4y_{\rm s}}{1+k}\right)
\frac{\delta\eta'}{\eta_{\rm b}}
-\frac{1-k}{1+k}\left(\frac{\delta \eta'}{\eta_{\rm b}}\right)^{2}=
\lim_{z\to z_{s}}\left[\frac{1}{2}\left(\frac{k}{V^{2}}-1\right)
\frac{\delta \eta''}{\eta_{\rm b}}\right],
\label{eq:higher_derivative_perfect_fluid}
\end{equation}
where the finite limit value on the right-hand side exists
because the left-hand side has a finite limit value. 
Employing the proof given in Appendix~\ref{sec:higher_derivative},
we can show that the limit value on the right-hand side is zero.
Then, we have (Eq.(4.15) of paper I)
\begin{equation}
\frac{\delta \dot{\eta'}}{\eta_{\rm b}}+\left(
-2\frac{1-k}{1+k}\frac{\eta'_{\rm b}}{\eta_{\rm b}}+\frac{5-3k-4y_{\rm s}}{1+k}\right)
\frac{\delta\eta'}{\eta_{\rm b}}
-\frac{1-k}{1+k}\left(\frac{\delta \eta'}{\eta_{\rm b}}\right)^{2}=0.
\label{eq:evolution_equation_perfect_fluid}
\end{equation}
With $k=1$, 
we find that the equation for full-order
perturbation at the sonic point is given by 
\begin{equation}
\frac{\delta \dot{\eta'}}{\eta_{\rm b}}+(1-2y_{\rm s})
\frac{\delta\eta'}{\eta_{\rm b}}=0.
\label{eq:kink_mode_equation_stiff_fluid}
\end{equation}
It should be noticed that the full-order analysis results in
a linear-order equation.
This equation is integrated as
\begin{equation}
\frac{\delta \eta'}{\eta_{\rm b}}=\mbox{const}\cdot e^{\alpha\tau},
\label{eq:kink_mode_solution_for_stiff_fluid}
\end{equation}
where
\begin{equation}
\alpha\equiv -(1-2y_{\rm s}).
\label{eq:growth_rate_for_stiff_fluid}
\end{equation}
Therefore, it is found that as $\tau$ increases, 
the discontinuity in $\eta'$ decays exponentially for $0<y_{\rm s}<1/2$
but grows exponentially for $1/2<y_{\rm s}$.
We should note that the parameter $y_{\rm s}$ can be 
identified with $a_{\sigma}$ by the discussion
given in Sec.~\ref{subsec:self-similar_stiff-fluid_solutions}.

This result for the stiff-fluid case 
is quite different from that for $0<k<1$ in the following respect.
For the latter case, the instability results in 
the divergence of $\delta \eta'$ for a finite value of $\tau$
because of its nonlinear growth.
However, for the former stiff-fluid case, the 
instability is much weaker and only exponential 
with respect to $\tau$.
This weakness of kink instability for the stiff-fluid case
may make the physical interpretation of instability 
rather subtle as is mentioned in paper I.
However, here we define instability by the 
exponential growth of discontinuity.

Then we find the following criterion: 
for early-time solutions,
solutions with a regular sonic point and $0<y_{\rm s}<1/2$
are stable against the kink mode, while 
those with a regular sonic point and $1/2<y_{\rm s}$
are unstable against the kink mode.
The situation is reversed for late-time solutions.
If the criterion for kink instability
should be related to the classification of sonic points,
we can give the following criterion:
as for early-time solutions,
nodal-point secondary-direction solutions
are stable, while
nodal-point primary-direction solutions
and 
saddle-point, repulsive-eigenvalue-direction solutions
are unstable. 
The situation is reversed for late-time solutions.
The above indicates that the stability criterion 
in terms of the properties of sonic points, 
which is obtained in paper I for $0<k<1$, 
also applies to the stiff-fluid case $k=1$.

\section{Correspondence between scalar-field and stiff-fluid solutions}
\label{sec:correspondence}
\subsection{Correspondence between a scalar field and a stiff fluid}
\label{subsec:correspondence_fluid_field}
We briefly review the correspondence between 
a scalar field and a stiff fluid.
The stress-energy tensor for the stiff fluid is written by
Eq.~(\ref{eq:stress-energy_tensor_of_stiff_fluid}).
We assume $\rho>0$.
Introducing a scalar field $\phi$ with a future-pointing timelike 
gradient such as
\begin{equation}
\nabla_{a}\phi=\sqrt{2\rho}u_{a},
\label{eq:timelikegradient}
\end{equation}
which is possible only for an irrotational velocity field,
we have
\begin{eqnarray}
\rho&=&-\frac{1}{2}\nabla_{c}\phi\nabla^{c}\phi, 
\label{eq:energy_density}\\
u_{a}&=&\frac{\nabla_{a}\phi}{\sqrt{-\nabla_{c}\phi\nabla^{c}\phi}},
\label{eq:four_velocity}
\end{eqnarray}
and it turns out that the stress-energy tensor is written by 
Eq.~(\ref{eq:stress-energy_tensor_of_scalar_field}), i.e.
that of a massless scalar field.
Therefore, the irrotational stiff fluid with positive energy density 
has an equivalent scalar field.
Actually, the correspondence is not one to one but
one stiff-fluid solution to two scalar-field solutions  
because $\tilde{\phi}=-\phi$ is also an equivalent solution
changing the relation (\ref{eq:timelikegradient}) as
\begin{equation}
-\nabla_{a}\tilde{\phi}=\sqrt{2\rho}u_{a}.
\label{eq:timelikegradientminus}
\end{equation}
Inversely, the scalar-field solution has an equivalent 
physical stiff fluid only if it has a timelike gradient.

\subsection{Correspondence between self-similar scalar-field solutions
and self-similar stiff-fluid solutions}
Here we obtain the correspondence between 
the self-similar scalar-field solutions
studied in Sec.~\ref{subsec:self-similar_scalar-field_solutions}
in the Bondi coordinates
and the self-similar stiff-fluid
solutions studied 
in Sec~\ref{subsec:self-similar_stiff-fluid_solutions} 
in the comoving coordinates.

In the Bondi coordinates~(\ref{eq:Bondi_coordinates}),
Eq.~(\ref{eq:energy_density}) yields
\begin{equation}
\rho=\frac{1}{2g}\phi_{,R}\left(2\phi_{,u}
-\bar{g}\phi_{,R}\right)=\frac{1}{2R^{2}}\frac{j}{g}[2x(j+\kappa)-\bar{g}j].
\label{eq:energydensitybondicoordinate}
\end{equation}
The relation between the time coordinates $t$ and $u$ is
given by the two equivalent expressions for the proper time 
element at the regular centre as
\begin{equation}
e^{\sigma/2}(-\infty)dt= \sqrt{g\bar{g}}(-\infty)du,
\end{equation}
where the arguments of the left-hand and right-hand sides are
$z$ and $X$, respectively. 
Integrating the above equation, we have
\begin{equation}
e^{\sigma/2}(-\infty)t= \sqrt{g\bar{g}}(-\infty)u.
\end{equation}
Then we can write the parameter $a_{\sigma}$
of stiff-fluid solutions in terms of the 
quantities of scalar-field solutions as
\begin{eqnarray*}
a_{\sigma}&=&\lim_{z\to -\infty}\eta e^{\sigma}e^{-2 z}, \nonumber \\
          &=&\lim_{r\to 0}8\pi \rho(t,r)r^{2} e^{\sigma}
	   \left(\frac{-t}{r}\right)^{2}, \nonumber \\
          &=&\lim_{R\to 0}8\pi \rho(u,R)R^{2} g\bar{g}
	   \left(\frac{-u}{R}\right)^{2}, \nonumber \\
          &=&\lim_{X\to -\infty}4\pi x^{-2}\bar{g}j[2x(j+\kappa)-\bar{g}j].
\end{eqnarray*}
Therefore, the relation between the self-similar stiff-fluid solution and 
scalar-field solution is given by 
\begin{equation}
a_{\sigma}=4\pi \kappa^{2},
\label{eq:correspondence_between_two_solutions1}
\end{equation}
where Eqs.~(\ref{eq:gbar0_g0_h0_j0}) and 
(\ref{eq:j'zero}) are used.

Moreover, we can explicitly show that the stiff-fluid solution 
which is equivalent to the scalar-field solution with regular centre
and regular similarity horizon satisfies 
the regularity condition at the sonic point.
The definition of the Misner-Sharp mass yields
\begin{equation}
m=\frac{R}{2}\left(1-\frac{\bar{g}}{g}\right).
\end{equation}
Then the function $y$ is written in terms of the quantities
associated with the scalar-field solution as
\begin{equation}
y=\frac{m}{4\pi\rho R^{3}} 
=\frac{g-\bar{g}}{4\pi j[2x(j+\kappa)-\bar{g}j]}.
\label{eq:y_in_Bondi_coordinates}
\end{equation}
Then we obtain 
\begin{equation}
y_{\rm s}=4\pi \kappa^{2},
\label{eq:correspondence_between_two_solutions2}
\end{equation}
at the sonic point, where Eqs.~(\ref{eq:regularity_similarity_horizon}) 
or (\ref{eq:regularity_similarity_horizon_kappa0}) are used.
Therefore, the regularity condition (\ref{eq:sonicpoint2})
at the sonic point for stiff-fluid solutions
is satisfied.

Equation~(\ref{eq:correspondence_between_two_solutions2}) 
is a crucial correspondence relation between
a self-similar scalar-field solution and  
self-similar stiff-fluid solution.
As expected, when there is the equivalence between 
a self-similar scalar-field solution and a self-similar 
stiff-fluid solution, the stability against the kink mode
coincides for both self-similar solutions,
as is seen from Eqs.~(\ref{eq:kink_mode_solution_for_scalar_field}), 
(\ref{eq:growth_rate_for_scalar_field}), 
(\ref{eq:kink_mode_solution_for_stiff_fluid}) and
(\ref{eq:growth_rate_for_stiff_fluid}).
The growth rates of the kink mode are also the same.
\section{Applications}
\subsection{Evans-Coleman stiff-fluid solution and critical behaviour}
\label{subsec:Evans-Coleman}
There are two important sequences of numerical self-similar 
solutions with analyticity
for a perfect fluid with $P=k\rho$, the one is the Larson-Penston 
solution and the other is the Evans-Coleman solution. 
The Larson-Penston solution in general relativity was
first discovered and discussed in the context of 
cosmic censorship by Ori and Piran~\cite{op1987},
and further analysis was done~\cite{op1990,fh1993,harada2001}.
The solution crosses a node along a secondary direction
for $0<k\alt 0.036$,
a degenerate node for $k\approx 0.036$, and
a node along a primary direction for $0.036\alt k<1/3$.   
However, the existence of the sequence of 
Larson-Penston solutions for a stiff fluid has not been demonstrated yet. 
Actually, in our preliminary numerical survey,
the existence of the Larson-Penston solution was confirmed
only for $0<k<1/3$.

The Evans-Coleman solution was first discovered
by Ori and Piran~\cite{op1990},  
identified with a critical solution by Evans and Coleman for $k=1/3$,
and shown to exist and identified 
with a critical solution 
for $0<k\le 1$ by several authors~\cite{maison1996kha1999,nc2000,bcgn2002}.
Carr {\it et al.}~\cite{ccgnu2000} indicated that 
the character of the sonic point for the Evans-Coleman 
solution changes as follows:
the solution crosses
a saddle along an attractive-eigenvalue direction for $0<k\alt 0.41$,
a node along a secondary direction for $0.41\alt k\alt 0.89$,
a degenerate node for $k\approx 0.89$,
and a node along a primary direction for $0.89\alt k <1$.

For the scalar-field system, 
Brady {\it et al.}~\cite{bcgn2002}
numerically solved the set of ordinary differential
equations (\ref{eq:g'})--(\ref{eq:j'}) 
and searched the value of $\kappa$ for which the solution 
has regular centre and analytic similarity horizon.
They found that there is such a solution 
with $4\pi \kappa^{2}\approx 0.577$.
We refer to this self-similar solution as the 
Brady-Choptuik-Gundlach-Neilsen (BCGN)
solution. They found that this solution 
has a single unstable mode which is analytic at the 
similarity horizon, and
that no physical stiff fluid can
reproduce this solution because the BCGN solution 
has a spacelike hypersurface $X=X_{1}(>X_{\rm s})$, 
on which the gradient of the scalar field is
null and beyond which it is spacelike. 
They mentioned that it is possible to consider another 
continuation beyond $X=X_{1}$ and then the solution 
can be regarded as a physical stiff-fluid solution.
Although this solution has some discontinuity at $X=X_{1}$,
it can be considered as a continuous limit of 
the Evans-Coleman perfect-fluid solution with $P=k\rho$
for $0.89\alt k<1$ (see Fig. 5 of~\cite{bcgn2002}).
Hence we refer to this solution as the Evans-Coleman
stiff-fluid solution. Neilsen and Choptuik~\cite{nc2000}'s 
numerical simulation suggests that this solution 
would be a critical solution of stiff-fluid gravitational 
collapse at the threshold of black hole formation
(see Fig. 2 of~\cite{bcgn2002}).

First we consider the stability of the Evans-Coleman stiff-fluid solution.
The parameter for this solution is given by
$a_{\sigma}=y_{\rm s}\approx 0.577$ using the correspondence relations 
(\ref{eq:correspondence_between_two_solutions1})
and 
(\ref{eq:correspondence_between_two_solutions2}).
Then, based on the analysis in 
Sec.~\ref{subsec:self-similar_stiff-fluid_solutions},
the sonic point of the Evans-Coleman stiff-fluid solution
is a node and
the solution crosses along a primary direction.
This is quite reasonable for a $k=1$ Evans-Coleman solution
since the Evans-Coleman solution crosses a node along a
primary direction for $k\to 1$.
From the stability criterion obtained in 
Sec.~\ref{subsec:kink_instability_stiff_fluid},
the Evans-Coleman stiff-fluid solution is unstable against the kink mode
because $y_{\rm s}\simeq 0.577>1/2$.
Actually, the stability against the kink mode does not change
whatever continuation one chooses beyond $X=X_{1}>X_{\rm s}$. 
The Evans-Coleman stiff-fluid 
solution has two unstable modes, 
one is analytic and the other is kink.
This implies that this solution cannot be a critical solution
of stiff-fluid gravitational collapse
if we allow initial data with sufficiently small discontinuity in the 
density gradient field.
It is discussed in paper I
why the kink instability did not affect
Neilsen and Choptuik~\cite{nc2000}'s fluid simulation.

\subsection{BCGN scalar-field solution and critical behaviour}
\label{subsec:BCGN}
Based on the analysis in Sec.~\ref{subsec:kink_instability_scalar_field}
the BCGN scalar-field solution is unstable against 
the kink mode because $4\pi \kappa^{2}\simeq 0.577>1/2$.
The BCGN scalar-field
solution has two unstable modes, 
one is analytic and the other is kink.
This implies that this solution cannot be a critical solution
of scalar-field gravitational collapse
if we allow initial data with sufficiently small discontinuity in the
second-order derivative of the scalar field.

Since this solution was found to have a single analytic 
unstable mode,
this solution is expected to act as an intermediate attractor
at least locally 
for some critical behaviour related to the 
asymptotic behaviour of gravitational collapse
from the point of view proposed by~\cite{kha1995}
if we restrict our attention to evolution
with $C^{\infty}$ functions for instance.
However, no numerical simulation has been reported so far, 
that suggests the existence of critical behaviour 
associated with this solution.
Brady {\it et al.}~\cite{bcgn2002} stated that the BCGN scalar-field 
solution is disqualified as a critical solution 
at the threshold of black hole
formation because it has an apparent horizon beyond the similarity
horizon. Since Choptuik~\cite{choptuik1993}'s numerical 
simulation breaks down at the formation of an 
apparent horizon because of the choice of 
coordinate system, it would be reasonable that the critical
behaviour associated with the BCGN scalar-field solution 
was not observed in his calculation. 
However, Hamad{\'e} and Stewart~\cite{hs1996}
did not mention the observation of critical behaviour 
associated with a continuously self-similar solution like
the BCGN scalar-field solution,
although their numerical simulation is 
based on the double-null formulation, which can
treat an apparent horizon without any coordinate singularity.

Here we would like to speculate about this puzzle, discrete 
self-similarity or continuous self-similarity.
Of course, we should note
the possibility that the numerical survey 
may not have been sufficiently complete yet.
If this possibility is true, 
further numerical investigation based on the double-null formulation
would reveal a {\em hidden} critical behaviour 
associated with the continuously 
self-similar solution, i.e. the BCGN scalar-field solution,
for which the critical exponent should be given by 
$\gamma=0.92\pm 0.02$. 
This critical behaviour associated with continuous self-similarity 
would be seen not at the black hole threshold but inside the 
apparent horizon.
In other words, it would be
hidden by an apparent horizon and therefore by an event horizon 
and could not be observed by a distant observer.  
We would have two distinct critical behaviours
for the same system, one is associated with
discrete self-similarity and the other is associated 
with continuous self-similarity.
However, we can suggest another possibility that the kink 
instability, which we have analysed here, may prevent the 
BCGN solution from involving critical behaviour.
The possibility depends on which class of initial data sets
we regard as physically realistic.
Mathematically one can avoid the kink instability by restricting 
oneself to solutions in $C^{\infty}$ class.
In numerical simulations, however, the differentiability condition 
may cause some subtle problems.
The above speculations can be proved or disproved by
further numerical investigations.
In each case, it is inferred that 
the qualitative properties of dynamical solutions in the spherical system 
of a massless scalar field would be 
much more complicated than usually supposed in the context 
of critical behavior.
\subsection{Exact self-similar solutions}
\label{subsec:exact_self-similar_solutions}
For the flat Friedmann solution, we find $4\pi \kappa^{2}=1/3$
as a scalar-field solution and $a_{\sigma}=y=1/3=\mbox{const}$ 
as a stiff-fluid solution. 
It is found that this solution is stable for early times (collapsing phase)
but unstable for late times (expanding phase).
This implies that there exists kink instability at the particle
horizon scale for the flat Friedmann scalar-field universe
and also for the flat Friedmann stiff-fluid universe.

The Roberts solutions are exact and general self-similar 
solutions for the scalar-field system with $\kappa=0$, 
which will be described in
Appendix~\ref{sec:roberts_solution}. 
According to the stability criterion
obtained here, these solutions do not suffer from kink instability
for early times while they do for late times.
Since this solution has a null gradient of the scalar field
at the similarity horizon, the stability criterion for self-similar 
stiff-fluid solutions is not applicable to this solution. 

We can find the static self-similar solution with a scalar field 
for $4\pi \kappa^{2}=1$, which is given by
\begin{equation}
\bar{g}=2x,\quad g=4x,\quad \bar{h}=\pm(4\pi)^{-1/2}\ln |x| +\mbox{const}.
\end{equation}
Also with a stiff fluid, we find the static self-similar solution as
\begin{equation}
M=\frac{1}{2}a_{\omega}^{1/4},\quad
S=a_{\omega}^{1/4},\quad
\eta=\frac{1}{2} a_{\omega}^{-1/2},\quad y=1,
\end{equation}
where the regularity condition (\ref{eq:sonicpoint2}) 
at the sonic point has been considered.
Apparently, this solution seems unstable for early times
but stable for late times.
We can show, however, that this solution is always unstable.
To demonstrate this explicitly, we change the coordinate system as
\begin{equation}
w\equiv -2\ln |u|+\ln R, 
\label{eq:utow}
\end{equation}
for the scalar-field solution, and
\begin{equation}
w\equiv \ln |t|, \quad R=a_{\omega}^{1/4}r,
\label{eq:ttow}
\end{equation}
for the stiff-fluid solution.
Then the line element for both cases 
is given in the static form as
\begin{equation}
ds^{2}=-2R^{2}dw^{2}+2 dR^{2}+R^{2}(d\theta+\sin\theta^{2}d\phi).
\label{eq:static_solution_static_chart}
\end{equation}
We should note that each of early-time and late-time solutions
corresponds to the whole spacetime of the static solution
through Eq. (\ref{eq:utow}) or (\ref{eq:ttow}).
Inversely, when Eq. (\ref{eq:static_solution_static_chart}) is given,
it can be regarded as an early-time solution 
using the coordinate transformation 
Eq. (\ref{eq:utow}) or (\ref{eq:ttow}).
In this solution, the similarity horizon condition $\bar{g}=2x$
or sonic point condition $V^{2}=1$
is satisfied everywhere.
It implies that we can insert a kink mode everywhere.
Then the evolution equation (\ref{eq:kink_mode_equation_scalar_field})
or (\ref{eq:kink_mode_equation_stiff_fluid}) for the kink mode 
perturbation holds along the radial null curve on which 
the discontinuity is inserted.
Therefore, when we insert a kink mode perturbation, which 
vanishes inside and is discontinuous at some radius,
it will blow up to infinity along the ingoing null curve
($w+\ln R=\mbox{const}$) as time proceeds.
In this sense, this solution is unstable.
The situation is somewhat different for a perfect fluid 
with $P=k\rho$ for $0<k<1$, which will be described in 
Appendix~\ref{sec:static_self-similar_perfect-fluid_solution}.

\subsection{Horizons}
For self-similar stiff-fluid solutions, a sonic point 
may be a black hole event horizon, particle horizon or 
Cauchy horizon, simultaneously.
Therefore, we can discuss the stability of these horizons.
These horizons appear for late-time solutions.
The black hole event horizon
is characterized by 
an anti-transsonic point for late-time solutions.
According to the result obtained here,
$y_{\rm s}>1$ must be satisfied on the event horizon.
Then the black hole event horizon corresponds to a saddle and
is stable against the kink mode.
The particle horizon and Cauchy horizon are characterized
by transsonic points for late-time solutions.
On these horizons we can have $0<y_{\rm s}<1/2$ or $1/2<y_{\rm s}<1$. 
These horizons correspond to nodes and 
are unstable if $0<y_{\rm s}<1/2$
while they are stable against the kink mode if $1/2<y_{\rm s}<1$.
The above also applies to similarity horizons in 
equivalent scalar-field solutions
using correspondence
relation~(\ref{eq:correspondence_between_two_solutions2}).

\subsection{Other applications}
Nonanalytic but regular self-similar stiff-fluid solutions are possible
only for $1/2<y_{\rm s}<1$. It is found that these solutions are
unstable for early times but stable for late times.
Anti-transsonic self-similar stiff-fluid 
solutions are possible only for $1<y_{\rm s}$.
These solutions are unstable for early times
but stable for late times.
Again, the above also applies to equivalent scalar-field solutions
using correspondence 
relation~(\ref{eq:correspondence_between_two_solutions2}).

\section{Summary}
We have investigated the stability of self-similar solutions 
with a scalar field and those with a stiff fluid in general relativity.
The kink instability, which we have considered here,
was studied in a Newtonian gas system~\cite{op1988} and
in the Einstein-perfect-fluid system with the equation of state 
$P=k\rho$ (paper I).
Since only the fluid system was considered in these previous 
works, it was not clear whether the kink instability 
is unique to fluid dynamics or not.
The present work, for the first time, has shown the 
existence of kink instability in self-similar solutions
of the Einstein-Klein-Gordon system.
The most intriguing feature is that the kink 
instability grows exponentially in terms of $T=-\ln|u|$
for the scalar-field system and also 
in terms of $\tau=-\ln |t|$ for the stiff-fluid system,
while it grows more rapidly and 
blows up to infinity at a finite moment before $t=0$
for the perfect-fluid system with $P=k\rho$ ($0<k<1$).
In other words, the kink instability results in 
the reduction of the rank of differentiability 
and the formation of a shock wave before the singularity formation
for a perfect fluid with $0<k<1$, 
while it does not for the scalar-field system and also 
for the stiff-fluid ($k=1$) system until the singularity forms.

These systems have recently attracted attention in the context of
critical phenomena in gravitational collapse.
The present result shows that both the 
BCGN scalar-field solution 
and 
Evans-Coleman stiff-fluid solution 
are 
unstable against the kink mode.
This implies that the latter solution, which was
identified with a critical solution in the stiff-fluid collapse by 
Neilsen and Choptuik~\cite{nc2000} and Brady {\it et al.}~\cite{bcgn2002},
cannot be a critical solution once we allow sufficiently small 
discontinuity in the density gradient field in the initial data sets;
nor can the BCGN scalar-field solution be a critical solution
once we allow sufficiently small 
discontinuity in the second-order derivative of the scalar field
in the initial data sets.
As another important application, we have shown the kink instability 
at the particle horizon of the flat Friedmann universe 
with a scalar field and with a stiff fluid,
while the flat Friedmann collapse solution is stable against this 
mode.

\acknowledgments
TH is grateful to B~J~Carr and R~Tavakol for helpful conversations.
The authors greatly appreciate important comments
by the anonymous referees, in particular
on the estimate of the higher-order derivative term in 
Eqs.~(\ref{eq:higher_derivative_scalar_field})
and (\ref{eq:higher_derivative_perfect_fluid}).
TH was supported by JSPS Postdoctoral Fellowship for Research Abroad.

\appendix
\section{Roberts solution}
\label{sec:roberts_solution}
In the Bondi coordinates~(\ref{eq:Bondi_coordinates}),
general solutions to Eqs.~(\ref{eq:g'})--(\ref{eq:j'})
with $\kappa=0$ are given by~\cite{roberts1989}
\begin{eqnarray}
g&=&\frac{x}{\sqrt{\sigma^{2}+x^{2}}}, \\
\bar{g}&=&\frac{\sqrt{\sigma^{2}+x^{2}}-2\sigma^{2}}{x}, \\
\phi&=&\frac{1}{4\sqrt{\pi}}\ln\frac{\sqrt{\sigma^{2}+x^{2}}-\sigma}
{\sqrt{\sigma^{2}+x^{2}}+\sigma},
\end{eqnarray}
where $x$ is defined by Eq.~(\ref{eq:xXT}), and 
the solutions are parametrized by $\sigma$ and referred to as 
Roberts solutions.
This solution has been discussed in the context of 
cosmic censorship~\cite{roberts1989} and critical 
behaviour~\cite{brady1994,ont1994}.

For $\sigma=0$ this solution reduces to the Minkowski spacetime
and we restrict $\sigma$ to $\sigma\ne 0$ below.
Then, these solutions admit no timelike Killing vector.
From Eq.~(\ref{eq:energy_density}), the
energy density of the equivalent stiff fluid is formally given by
\begin{equation}
\rho=\frac{1}{8\pi R^{2}}\frac{\sigma^{2}(2\sqrt{\sigma^{2}+x^{2}}-1)}
{x^{2}}.
\end{equation}
We can see that the solution has a spacetime singularity at $x=0$.
If and only if $0<|\sigma|<1/2$ this solution has a region around 
the centre given by
\begin{equation}
0<x<x_{1}=\sqrt{\frac{1}{4}-\sigma^{2}},
\end{equation}
where the gradient of the scalar field is spacelike.
At $x=x_{1}$ the solution has a null gradient of the scalar field. 
If and only if $0< |\sigma| <1/2$, the solution has a 
similarity horizon at 
\begin{equation}
x=x_{\rm s}=\sqrt{\frac{1}{4}-\sigma^{2}}.
\end{equation}
Therefore, for $0<|\sigma|<1/2$, the hypersurface $x=x_{1}$ where
the gradient of the scalar field is null coincides with
the similarity horizon $x=x_{\rm s}$.
This implies that the stability analysis for self-similar 
stiff-fluid solutions
developed in Sec.~\ref{subsec:kink_instability_stiff_fluid} 
cannot be used, although that for 
self-similar scalar-field solution developed 
in Sec.~\ref{subsec:kink_instability_scalar_field}
is still applicable.

See~\cite{frolov1997} for spherical and nonspherical perturbation
with analyticity on the Roberts solution. 

\section{Higher-order derivative term}
\label{sec:higher_derivative}
Let $f$ be continuous on $[0,\infty)$ and 
continuously differentiable on $(0,\infty)$ and have
the limit value $\lim_{x\to +0} x f'(x)$,
where the prime denotes the differentiation with respect to the argument.
When we define $g$ by $g(x)\equiv xf(x)$,
$g$ is continuous on $[0,\infty)$ with $g(0)=0$
and continuously differentiable on $(0,\infty)$
and have the limit value $\lim_{h\to +0} g(h)/h=f(0)$.
Then, from the mean value theorem, for any 
$a>0$, there exists $\xi $ ($0<\xi <a$), such that
\begin{equation}
\frac{g(a)}{a}=g'(\xi).
\end{equation}
In terms of $f$, we have 
\begin{equation}
f(a)=\xi f'(\xi)+f(\xi).
\end{equation}
When we take the limit $a\to +0$, $\xi$ also goes to zero.
Since the left-hand side and the second term on the right-hand side 
approach the same limit value $f(0)$, the first term
on the right-hand side goes to zero in this limit.
This implies
\begin{equation}
\lim_{x\to +0}xf'(x)=0,
\end{equation}
because we have assumed the existence of the above limit.

Applying the above proof, we can show that the right-hand side 
vanishes in
Eq.~(\ref{eq:higher_derivative_scalar_field}) for the scalar-field system,
where $f$ corresponds to $\delta j'$, 
and in Eq.~(\ref{eq:higher_derivative_perfect_fluid}) 
for the perfect-fluid system,
where $f$ corresponds to $\delta \eta '$.

\section{static self-similar perfect-fluid solution}
\label{sec:static_self-similar_perfect-fluid_solution}
We briefly review the static self-similar solution 
for a perfect fluid with the equation of state $P=k\rho$
for $0<k<1$. The case of $k=1$ is described 
in Sec.~\ref{subsec:exact_self-similar_solutions}.
We use $t$ and $r$ as the coordinates given by Eq.~(\ref{eq:diagonal}).
For $0<k<1$, the line element 
in the static self-similar solution is written as
\begin{equation}
ds^{2}=-\alpha_{k}^{2} R^{\frac{4k}{1+k}}dw^{2}
+\beta_{k}^{2} dR^{2}+R^{2}(d\theta^{2}+\sin^{2}
\theta d\phi),
\end{equation}
after the following coordinate transformation
\begin{eqnarray}
w&=&
\left\{
\begin{array}{ll}
\displaystyle{-\frac{1+k}{1-k}(-t)^{\frac{1-k}{1+k}}+w_{0}}& \quad \mbox{for}
\quad  t<0, \\
\displaystyle{\frac{1+k}{1-k}t^{\frac{1-k}{1+k}}+w_{0}}& \quad \mbox{for}
\quad t>0, \\
\end{array}
\right.\\
R&=&\gamma_{k}r,
\end{eqnarray}
where $\alpha_{k}$, $\beta_{k}$ and $\gamma_{k}$ 
are some positive constants which can depend
on $a_{\sigma}$, $a_{\omega}$ and $k$, 
and $w_{0}$ is an arbitrary constant.
In the above, $w$ increases from $-\infty$ to $+\infty$
as $t$ increases from $-\infty$ to $+\infty$.
Therefore, the static self-similar solution consists of both 
the early-time solution and the late-time solution.
We can identify an arbitrary value of $w$ with $t=0$ because
of the arbitrary constant $w_{0}$.
Therefore any value of the time coordinate $w$ can be 
regarded as in early time and also in late time.
This is essential to the instability of the static 
self-similar solution as is discussed in paper I.


\begin{thebibliography}{99}
\bibitem{ct1971}
Cahill~M~E and Taub~A~H 1971
{\it Comm. Math. Phys.} {\bf 21} 1
\bibitem{gnu1998}
Goliath~M, Nilsson~U and Uggla~C
1998 {\it Class. Quantum Grav.} {\bf 15} 167;\\ 
Goliath~M, Nilsson~U and Uggla~C
1998 {\it Class. Quantum Grav.} {\bf 15} 2841
\bibitem{cc2000}
  Carr~B~J and Coley~A~A
  2000 {\it Phys. Rev.} D
  {\bf 62} 044023;\\
  Carr~B~J and Coley~A~A
  2000 {\it Class. Quantum Grav.} 
  {\bf 17} 4339
\bibitem{ccgnu2000}
Carr~B~J, Coley~A~A, Goliath~M, Nilsson~U and Uggla~C
2000
{\it Phys. Rev.} D {\bf 61} 081502;\\
Carr~B~J, Coley~A~A, Goliath~M, Nilsson~U and Uggla~C
2001 
{\it Class. Quantum Grav.} {\bf 18} 303
\bibitem{we1997}
  Wainwright~J and Ellis~G~F~R
  {\it ``Dynamical Systems in Cosmology''}
  (Cambridge University Press, Cambridge, England, 1997)
\bibitem{coley1999}
  Coley~A~A 1999
  gr-qc/9910074
\bibitem{ch1974}
  Carr~B~J and Hawking~S~W
  1974 {\it Mon. Not. R. Astr. Soc.} {\bf 168} 399
\bibitem{lcf1976}
  Lin~D~N~C, Carr~B~J and Fall~S~M
  1976 {\it Mon. Not. R. Astr. Soc.} {\bf 177} 51
\bibitem{bh1978}
  Bicknell~G~V and Henriksen~R~N
  1978 {\it ApJ}
  {\bf 219} 1043;\\
  Bicknell~G~V and Henriksen~R~N
  1978 {\it ApJ} {\bf 225} 237
\bibitem{op1987}
  Ori~A and Piran~T
  1987 {\it Phys. Rev. Lett.}
  {\bf 59} 2137;\\
  Ori~A and Piran~T
  1988 {\it Gen. Rel. Grav.}
  {\bf 20} 7
\bibitem{op1990}
  Ori~A and Piran~T
  1990 {\it Phys. Rev.} D
  {\bf 42} 1068
\bibitem{fh1993}
  Foglizzo~T and Henriksen~R~N
  1993 {\it Phys. Rev.} D {\bf 48} 4645
\bibitem{cc1999}
  Carr~B~J and Coley~A~A
  1999 {\it Class. Quantum Grav.} {\bf 16} R31-R71
\bibitem{carr1993}
  Carr~B~J 1993 unpublished
\bibitem{hm2001}
Harada~T and Maeda~H, 
2001 {\it Phys. Rev.} D {\bf 63} 084022
\bibitem{harada1998}
Harada~T
1998 {\it Phys. Rev.} D {\bf 58} 104015
\bibitem{choptuik1993}
  Choptuik~M~W
  1993 {\it Phys. Rev. Lett.}
  {\bf 70} 9
\bibitem{hs1996}
  Hamad{\'e}~S~H and Stewart~J~M
  1996 {\it Class. Quantum. Grav.} {\bf 13} 497
\bibitem{ec1994}
  Evans~C~R and Coleman~J~S
  1994
  {\it Phys. Rev. Lett.}
  {\bf 72} 1782
\bibitem{kha1995}
  Koike~T, Hara~T and Adachi~S
  1995 
  {\it Phys. Rev. Lett.}
  {\bf 74} 5170
\bibitem{gundlach19951997}
  Gundlach~C
  1995
  {\it Phys. Rev. Lett.}
  {\bf 75} 3214;\\
  Gundlach~C
  1997
  {\it Phys. Rev. D}
  {\bf 55} 695
\bibitem{maison1996kha1999}
  Maison~D
  1996 {\it Phys. Lett.} B
  {\bf 366} 82;\\
Koike~T, Hara~T and Adachi~S
1999 {\it Phys. Rev.} D {\bf 59} 104008
\bibitem{nc2000}
Neilsen~D~W and Choptuik~M~W
2000 {\it Class. Quantum Grav.} {\bf 17} 733;\\
Neilsen~D~W and Choptuik~M~W
2000 {\it Class. Quantum Grav.} {\bf 17} 761
\bibitem{mh2001}
Maeda~H and Harada~T
  2001 {\it Phys. Rev.} D {\bf 64} 124024;\\
  Harada~T, Maeda~H and Semelin~B
  2003 {\it Phys. Rev. } D {\bf 67} 084003
\bibitem{madsen1988cg1999}
  Madsen~M~S
  1988 {\it Class. Quantum Grav.} {\bf 5} 627;\\
  Carr~B~J and Goymer~C~A
  1999 {\it Prog. Theor. Phys. Suppl.} {\bf 136} 321
\bibitem{bcgn2002}
  Brady~P~R, Choptuik~M~W, Gundlach~C and Neilsen~D~W
  2002 {\it Class. Quantum Grav.} {\bf 19} 6359
\bibitem{gundlach2003}
  Gundlach~C 
  2003 {\it Phys. Rep.} {\bf 376} 339
\bibitem{he1973}
  Hawking~S~W and Ellis~G~F~R
  {\it ``The large scale structure of space-time''}
  (Cambridge University Press, Cambridge, England, 1973)
\bibitem{op1988}
Ori~A and Piran~T 
1988
{\it MNRAS} {\bf 234} 821
\bibitem{harada2001}
  Harada~T
  2001 {\it Class. Quantum Grav.} {\bf 18} 4549
\bibitem{wald1983}
  Wald~R~M
  {\it ``General Relativity''} (University of Chicago Press,
	Chicago, United States of America, 1983)
\bibitem{christodoulou1994}
  Christodoulou~D
  1994 {\it Annals of Math} {\bf 140} 607
\bibitem{brady1995}
  Brady~P
  1995 {\it Phys. Rev.} D {\bf 51} 4168
\bibitem{roberts1989}
  Roberts~M~D
  1989 {\it Gen. Rel. Grav.} {\bf 21} 907
\bibitem{ms1964}
Misner~C~W and Sharp~D~H
1964 {\it Phys. Rev.} D {\bf 136} 571
\bibitem{footnote3}
Equation (3.32) of paper I contains a trivial typo.
\bibitem{footnote1}
The existence of the finite limit value for 
$\delta \dot{\eta}'$ is not 
explicitly assumed in paper I.
\bibitem{footnote2}
Equation~(\ref{eq:higher_derivative_perfect_fluid}) 
corresponds to Eq.~(4.15) in paper I, where 
the right-hand side is ignored without caution.
\bibitem{brady1994}
  Brady~P
  1994 {\it Class. Quantum Grav.} {\bf 11} 1255
\bibitem{ont1994}
  Oshiro~Y, Nakamura~K and Tomimatsu~A 
  1994 {\it Prog. Theor. Phys.} {\bf 91} 1265 
\bibitem{frolov1997}
  Frolov~A~V
  1997 {\it Phys. Rev.} D {\bf 56} 6433;\\
  Frolov~A~V
  1999 {\it Phys. Rev.} D {\bf 59} 104011;\\
  Frolov~A~V
  2000 {\it Phys. Rev.} D {\bf 61} 084006
\end{thebibliography}
\end{document}